# Hamamatsu PMT R7056 Study for the Extinction Monitoring System of the Mu2e Experiment at Fermilab

S. Boi, A. Dyshkant, D. Hedin, E. Johnson, E. Prebys, P. Rubinov

*Abstract*–**The Mu2e experiment at Fermilab proposes to search for the coherent neutrino-less conversion of muons to electrons in the presence of a nucleus. The experimental signature for an aluminum target is an isolated 105 MeV electron exiting the stopping target no earlier than ~700 ns after the pulse of proton beam hits the production target. Any protons that hit the production target in between the pulses can lead to fake conversion electrons during the measurement period. We define the beam extinction as the ratio of the number of protons striking the production target between pulses to the number striking the target during the pulses. It has been established that an extinction of approximately $10^{-10}$ is required to reduce the backgrounds to an acceptable level. It would be desirable to measure the extinction of the beam coming out of the accelerator in a minute or less. Studies for the fast extinction monitor based on Hamamatsu PMT R7056 is the subject of this presentation.**

## I. Introduction

Extended versions of the Standard Model (SM) predict a small rate of neutrino-less muon to electron conversion. If verified, charged lepton flavor violation (CLFV) would point to new physics beyond the SM. With sensitivity of four orders of magnitude better than previous experiment, the Mu2e experiment at Fermilab [1] will search for the coherent neutrino-less conversion of muon to electron CLFV processes in the field of an atomic nucleus. The Mu2e experiment consists of three consecutive large superconductive solenoids:
the production solenoid with the production target inside for a primary proton beam, where secondary particles are produced and focused;
the transport solenoid to transfer the beam of low energy muons through the volume with rotatable collimators for positive or negative particles;
and the detector solenoid enclosing the aluminum muon stopping target, particle tracker, and an electromagnetic calorimeter for the momentum measurements, identification of charged particles, and a fast triggering.
Mu2e also has a cosmic ray veto system, a Germanium stopping target monitor, and a proton beam extinction monitors to measure the fraction of out of time primary pulse protons which can cause a false conversion signal.

## II. Pulsed Beam and Background Supression

For the Mu2e experiment it is crucial to minimize the beam-induced backgrounds and uses proton pulses with the search window about 670 ns after, when backgrounds from the proton pulse have died away, and would continue until the next proton bunch. The measurement period of ~700 - 1700 ns after injection matches the 0.88 ms life time of muonic Al.
Backgrounds can also be produced by protons hitting the production target during or slightly before the search window. Suppression of prompt backgrounds requires a pulsed proton beam, in which the ratio of the amount of protons between pulses to the amount of protons contained in a pulse is less than $10^{-10}$. This ratio is defined as the beam extinction.
For the Mu2e experiment batches of 8 GeV protons will be produced in the Fermilab Booster and transported to the Recycler Ring where they will be re-bunched into four 2.5 MHz bunches. One of four bunches will be kicked out of the Recycler and transported/injected into the Delivery Ring. The bunch will be transferred from the Delivery Ring by a slow resonant extraction through beam line to the production target. An 8 GeV primary beam has about 31 million protons per pulse, 250 ns length, and a period of about 1695 ns. It produces bunches of muons that are transported to and stopped in the Al target.

### A. *About Extinction*

The required beam extinction will be achieved in two steps. First, the technique for generating the required bunch structure will naturally lead to a high level of extinction ~$10^{-5}$; and second, the beam line from the Delivery Ring to the production target has a set of oscillating dipoles (AC) to sweep out-of-time protons into a system of collimators. This should achieve an additional extinction of $10^{-7}$ or better.

Manuscript received November 6th, 2015. This work was supported in part by the US NSF, the US DoE, and was operated in part by Fermi Research Alliance, LLC under Contract No. De-AC02-07CH11359 with the United States Department of Energy

S. Boi, A. Dyshkant, D. Hedin, E. Johnson are with Northern Illinois University, De Kalb, IL 60115 USA (telephone: 815-753-1717, e-mail: dyshkant@nicadd.niu.edu).

E. Prebys, P. Rubinov are with Fermi National Accelerator Laboratory, Batavia, IL 60510 USA.

..

The Mu2e proton beam cycle and the delayed search window allows for the effective elimination of prompt backgrounds when the number of protons between pulses is suppressed to the required level.

### *B. Extinction Measurements*

Direct extinction measurement is hard because of a high rate. So, the Mu2e focuses on measuring beam scattered particles using a detector with good time resolution and small effective acceptance. Then over many bunches, a statistical picture can be built of the out of time population

There are two time scales in terms of an extinction measurement. To avoid long running periods with unexpected anomalies about an hour time scale is preferable for the high precision about $10^{-10}$ measurements.

Because there could be different subtle problems with the extinction in the Delivery Ring, it is important to monitor for potential failures of the beam delivery system on a much shorter time scale with the precision of about $10^{-5}$ or so.

For the high precision measurement the experiment has a dedicated production target extinction monitor. It detects particles with average momentum of 4.2 GeV/c and a time resolution of about 25 ns to separate particles produced "in time" with the proton bunches from those produced between bunches. It consist of a permanent magnet combined with entry and exit collimators. To isolate a sample of charged particles with well define direction, six planes of 4 cm x 4 cm pixel detectors will be used to reconstruct tracks that are closely align with the axis of the exit collimator.

To measure the level of extinction in the beam from the Delivery Ring on a faster time scale, a second monitor can be placed upstream of the AC dipole. The idea is to use a small obstruction in the beam path to create scattering that is detected by an external detector [2]. The key requirements to such a detector are a time resolution that is short compare to the nominal proton pulse length and a low fake rate. Similar techniques have been used to measure a beam halo [3].

### *C. Upstream Extinction Monitor*

Our technique is based on a telescope of four Cherenkov counters to register beam particles scattered off of a thin (5 μm Ti) foil installed in the beam line. The telescope has a limited acceptance tuned in such a way that scattering from the in-time protons will not cause saturation. Particle detection will be accomplished by four such telescopes. The telescopes are located outside of the beam line at about 2 m downstream of the foil and position in a straight line to its beam point intersection.

The Cherenkov light is read out by a PMT. The PMTs outputs will be connected to the waveform digitizers that will allow the time structure of the Cherenkov light to be analyzed in detail. Data acquisition will be synchronized to the 2.5 MHz RF clock. A μTCA crate will provide the framework for the data acquisition. A beam time profile will be built by integrating over many bunches. The general issues of such measurements are the need of large dynamic range, ability of the photo detector to withstand high rates, good time resolution, and no or low fake responses like, for example, correlated after pulses [4, 5] in the photomultiplier tubes used in coincidence.

The main telescope unit is a Cherenkov counter. It consists of a radiator and a photodetector. The following Cherenkov radiators have been tested with cosmic rays: fused quartz GE type 021, UV transparent PMMA [7], and Cherenkov plastic EJ-299-15 [7]. As a photodetector we tested Hamamatsu PMT R7056 and FEU-115M. Because those PMTs have similar dimensions, the FEU-115M housing [8] was employed for R7056. The R7056 has an advantage in being able to register the Cherenkov light because of the UV glass window.

## III. Test Measurements

### *A. PMT Tests for After Pulses*

The R7056s with the Hamamatsu (A) socket assembly E2624-14 were tested with cosmic rays at high voltage about 1.45 kV. The setup consisted of three Cherenkov counters. The radiators have rectangular shape with dimensions about 100 mm length, 27 mm width, and up to 12.7 mm thickness. Each radiator had orientation of the largest surface in the horizontal plane and assembled vertically on the top of each other, like a projection tower. The distance between the top and bottom counters was about 10 cm.

The top and the bottom counters coincidence was used as a trigger to a scope. The average trigger rate was about 3 counts per minute. The data acquisition system was the scope based with a LabVIEW program. The PMT outputs were then digitized by the scope. The true and after pulses were recognized as a minimal response in the appropriate time region. The after pulses produced by a photomultiplier tube were observed [4, 5] at about 20% relative frequency within a 1.8 μs time window after the true pulse.

In conjunction with the x10 PM amplifier LeCroy model 612A the high voltage was reduced to about 1 kV and the test was performed using a Hamamatsu (B) socket assembly. The after pulses were dropped to about few per cent relative frequency compare to the true pulses (see Fig. 1) due to both the lower high voltage applied to the PMT and also the lower voltage between the photocathode and the first dynode [4]. To reduce voltage at the first stage of PMT even more a custom made tapered voltage divider was used.

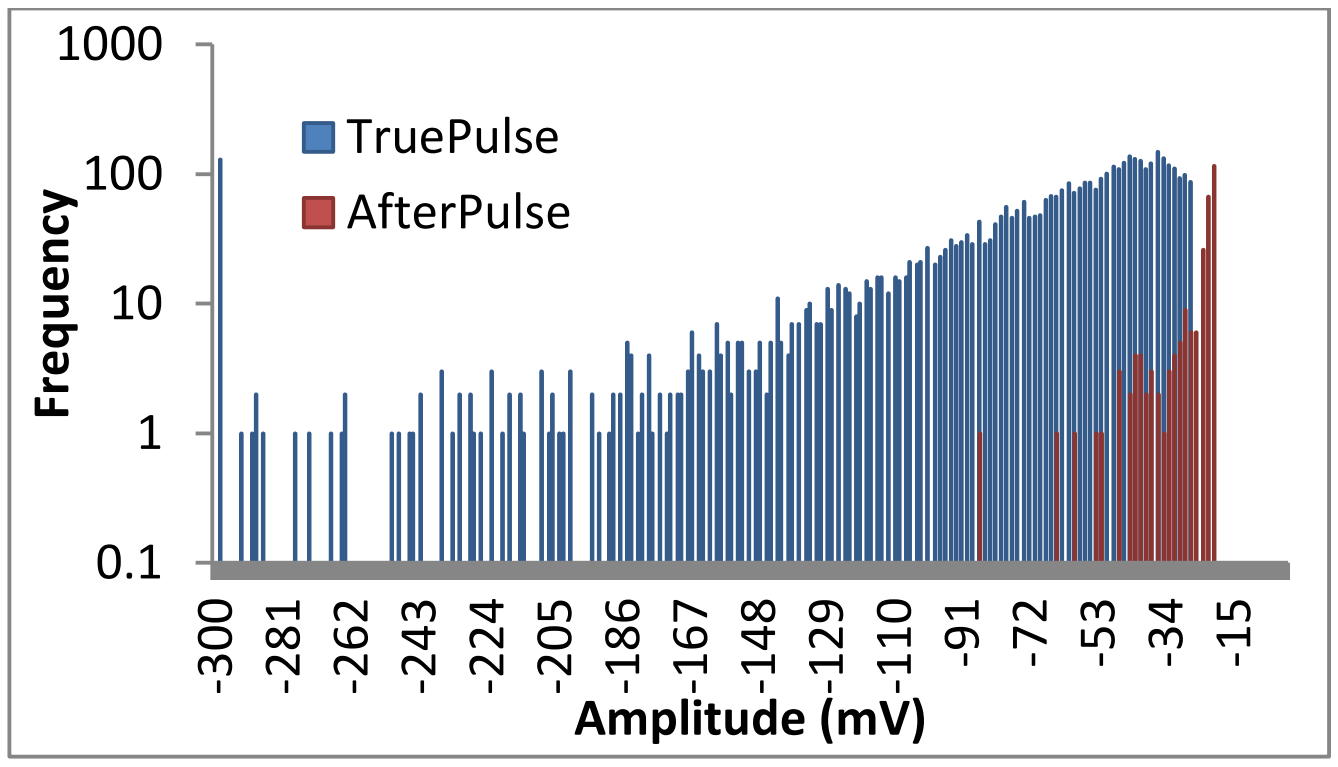

Fig. 1. Frequency of "True" Pulse and "After" Pulse Amplitudes

### *B. PMT Tests for High Rate*

To look at a PMT response at high frequencies, the PMT R7056 was instrumented with the LED5-UV-400-30 and, by modification of the illumination, tested at the large (about 500mV) and the low (about 10 mV) PMT output pulses in the frequency range from about 0.07 to 20 (30) MHz. The restriction of output pulses amplitude was observed because of a low value of the divider current (about 0.2 mA at 1 kV). At the low illumination the amplitude of the output pulses was very sensitive to a frequency and to an average value of the anode current. At about 2 μA average value of the anode current (at low illumination) the PMT was responding up to 20 MHz frequency of pulses from the generator to the LED. The observations in a pulse mode point out on importance of the average anode current level and a ratio of average anode current to the voltage divider current as well [9, 10]. To mimic a high rate the laboratory study were performed at some level of permanent light from LED presented and the voltage drop on the anode load and the voltage at the last dynode measured. The level of permanent light from the LED was modified by varying the DCV applied to the LED.

### *C. Custom Made Voltage Dividers*

The PMT factory anode load needs to be modified and a lower value is preferable [9, 10]. At a high rate the anode current will be large. If an anode load is low (about 50 Ω) then at a large anode current this will cause a few millivolts voltage reduction at the last PMT stage. In addition, the 50 Ω anode load will remove possible output pulse reflections.

Custom made tapered voltage dividers with the divider current about 0.25 mA, 2.5 mA, and 10 mA at 1kV were made and laboratory tested (Fig. 2).

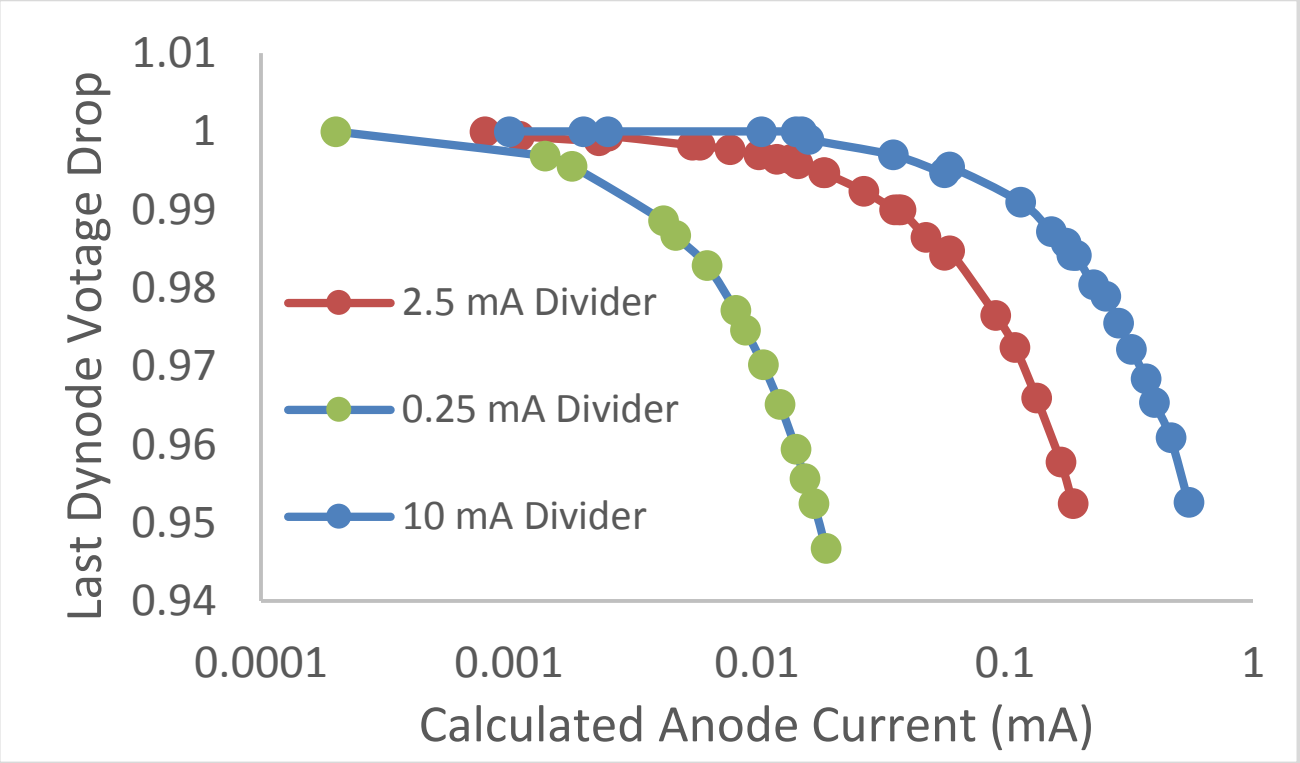

Fig. 2. Divider current effects on relative last dynode voltage drops at different anode currents

### *D. PMT-Radiator Optical Interface Tests*

A silicon optical grease and a silicon soft pad can improve the optical connection between the Cherenkov radiator and the PMT input window (see Fig. 3). But those silicon materials have absorption at shorter wavelengths. The UV-absorption edge is at about 300 nm and is sensitive to the thickness [7]. For those reasons (increase from reduced reflections but decrease from absorption) the interface between the radiator and the PMT input window was tested with and without a 1.5 mm thick soft silicon pad (see Fig.4).

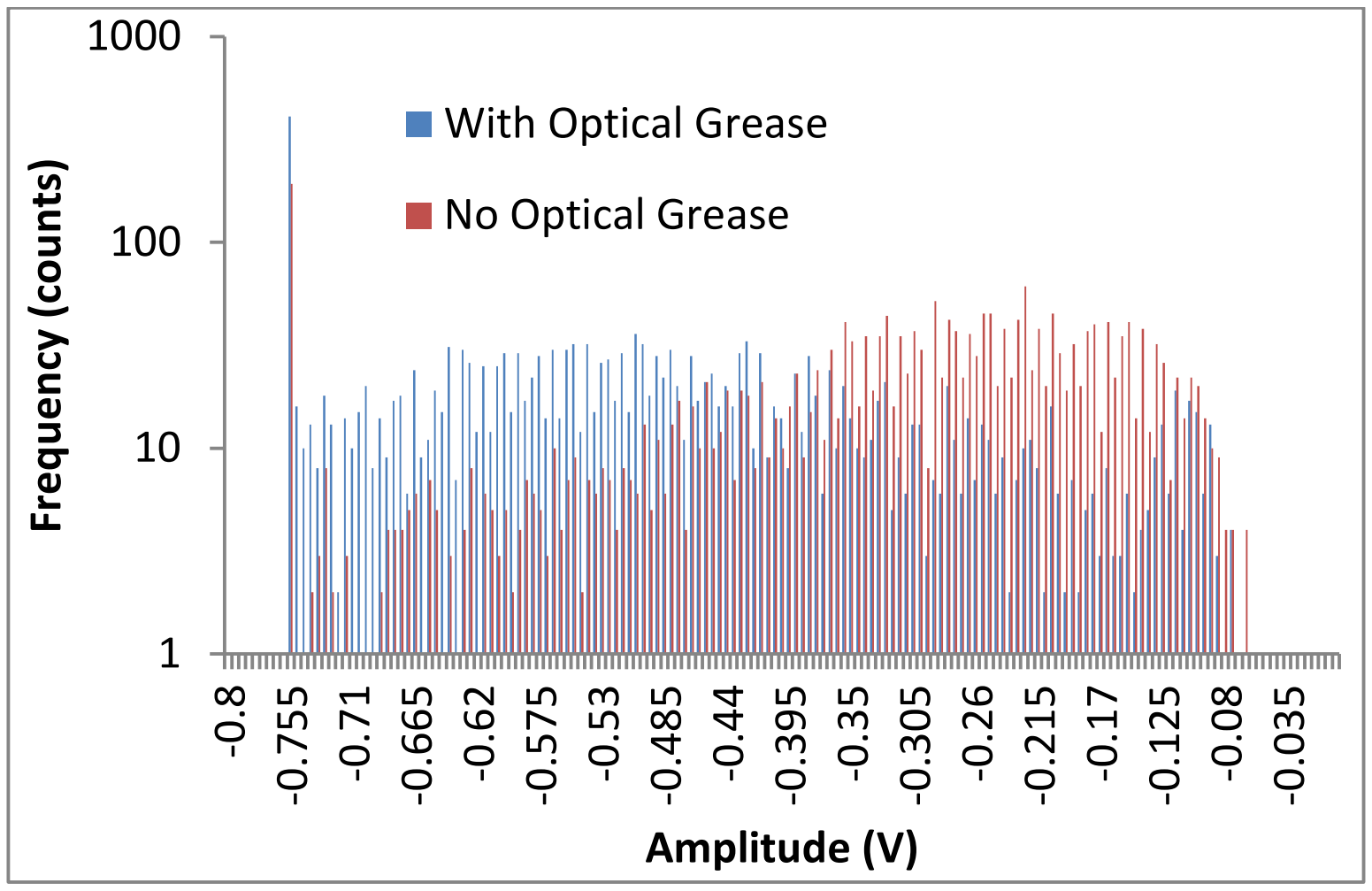

Fig. 3. Frequency of output pulse amplitudes from the Hamamatsu PMT R7056 connected to the quartz radiator

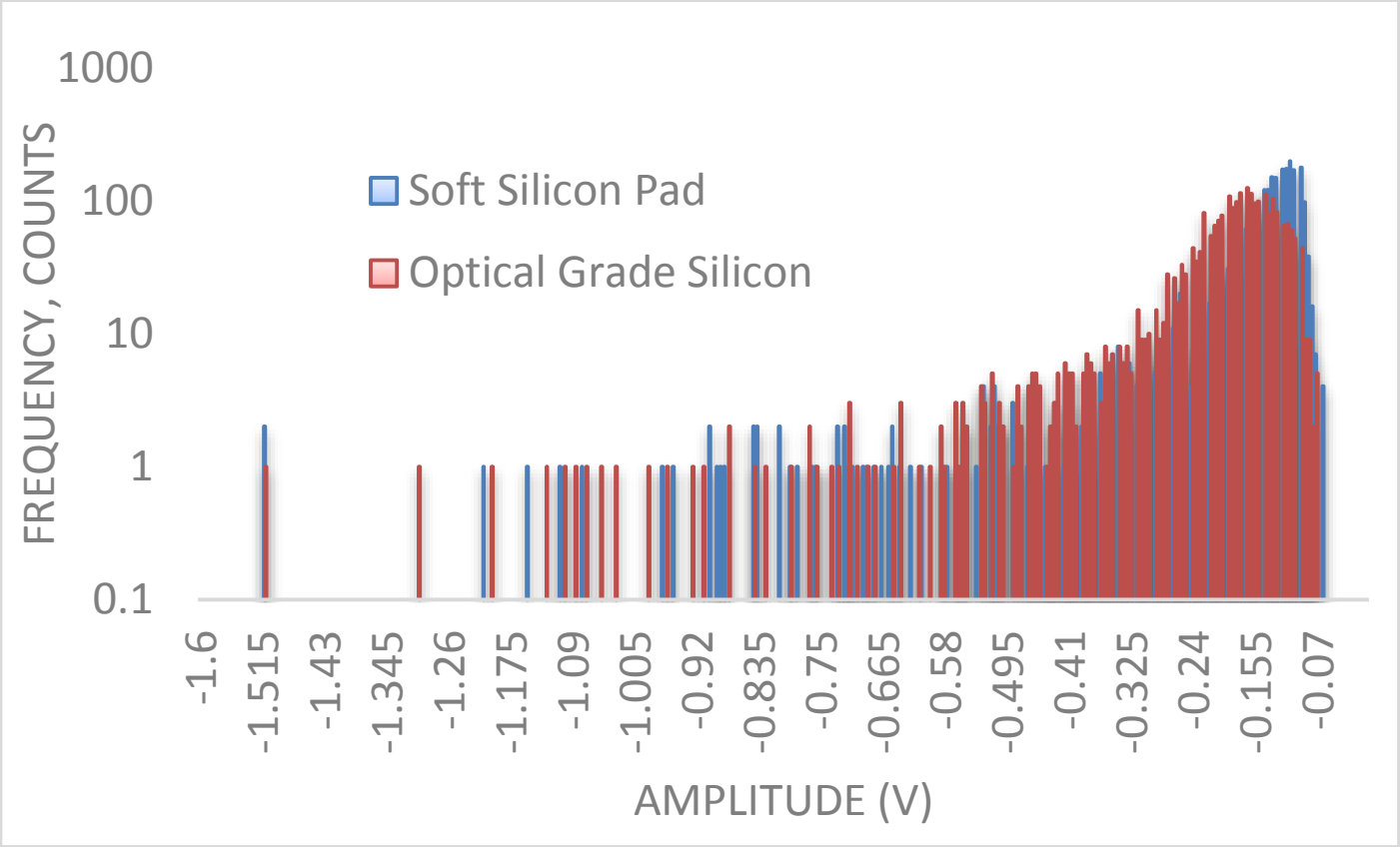

Fig. 4. Frequency of output pulse amplitudes for a soft silicon pad or an optical grease between the quartz radiator and the PMT R7056

## IV. Summary

For the Mu2e extinction measurements the Hamamatsu PMT R7056 is preferably used at a high voltage of about 1 kV with a custom made tapered voltage divider in conjunction with about x10 amplification.

To avoid or further reduce occurrence of after pulses the relative voltage between the photocathode and the first dynode was decreased by a factor of two.

At high rate the low anode load of 50 Ω was preferable.

A safe soft silicon pad interface between the radiator and the PMT window absorbs some Cherenkov light.

## Acknowledgment

The authors would like to thank the NIU engineer and technicians who contributed to the construction of the prototypes Mike Figora, Aaron Sturtz and William Vickers for preparation of the apparatus.